\newcommand{\bee}{\begin{equation}}
\newcommand{\ene}{\end{equation}}

\documentclass[10pt,pre,twocolumn,amsmath]{revtex4}
\usepackage[pdftex]{graphicx} 
\usepackage{color}

\begin{document}
\draft
\title{Plasma double layer development during high power EUV exposure}

\author{Manis Chaudhuri$^{1}$\footnote{Corresponding author: manis.chaudhuri@asml.com}, Pavel Krainov$^2$, Dmitry Astakhov$^2$, Andrei M. Yakunin$^1$, and Mark van de Kerkhof$^{1,3}$}
\affiliation{$^1$ASML The Netherlands B.V., P.O.Box 324, 5500AH Veldhoven, The Netherlands}
\affiliation{$^2$ISTEQ B.V., Eindhoven, The Netherlands}
\affiliation{$^3$Department of Applied Physics, Eindhoven University of Technology, PO Box 513, 5600 MB Eindhoven, The Netherlands}

\date{\today}

\begin{abstract}

The development of electrostatic plasma double layer (DL) at the boundary of Extreme Ultra-Violet (EUV) exposed and un-exposed region in the bulk volume has been confirmed by 3DPIC (Particle-In-Cell) simulations in the context of fast transient high power EUV exposures. It is found that the DL exists only for short time scale during EUV-ON time period ($\sim 70{\rm ns}$) and disappears soon after EUV is OFF. Such DL fingerprint appears above a certain critical value of EUV beam energy ($\sim 0.1 {\rm mJ}$) and it transforms from weak-to-strong DL with further increase of EUV power.   

\end{abstract}

\maketitle

Plasma double layers (DL) are self-organized, non-neutral and nonlinear structures consisting of adjacent regions of net positive and negative space charge that sustain a localized electrostatic potential drop within an otherwise quasi-neutral plasma~\cite{Langmuir1929,Alfven_book1981,Hershkowitz1985,Alfven1986}. Early laboratory and theoretical studies established DL as self-consistent solutions of the plasma kinetic equations, capable of maintaining electric fields over only a few Debye lengths while mediating significant energy transfer between fields and particles~\cite{Sagdeev1966,Block1978,Schamel1986,Raadu1989}. They have been observed in different laboratory plasmas such as mercury discharge~\cite{Torven1971,Strangeby1973}, Q-machines~\cite{Sato1976,Sato1982}, double plasma (DP) devices~\cite{Quon1976,Sekar1985}, triple plasma (TP) devices~\cite{Coakley1978}, helicon devices~\cite{Hairapetian1990,Charles2003} to laser pellet ablation plasmas~\cite{Hora1984}. Such laboratory experiments have demonstrated that DL can form spontaneously as current-driven DL (CDDL)~\cite{Sato1980,Sato1981,Cartier.Marlino1984,Hershkowitz2005} or current-free DL (CFDL)~\cite{Perkins1981,Hatakeyama1983,Hairapetian1988,Boswell2006,Liberman2006,Chen2006,Charles2007,Takahashi2007,Ahedo2009,Thakur2009,Takahashi2011}. Observations in space plasmas significantly broadened the relevance of DL physics such as discrete auroral arcs~\cite{Mozer1997,Ergun1998,Andersson2002}, the solar wind~\cite{Carlqvist1982} and in planetary magnetospheres~\cite{Carlqvist1982,Temerin1982}. These findings firmly established DL as a natural plasma phenomenon operating across many orders of magnitude in scale~\cite{Hultqvist1971,Torven1976,Stenzel1982,Yamamoto1985,Williams1986,Lennartsson1987,Lindberg1988,Raadu1988}.
\begin{figure}[h]
\includegraphics[width=0.95\linewidth]{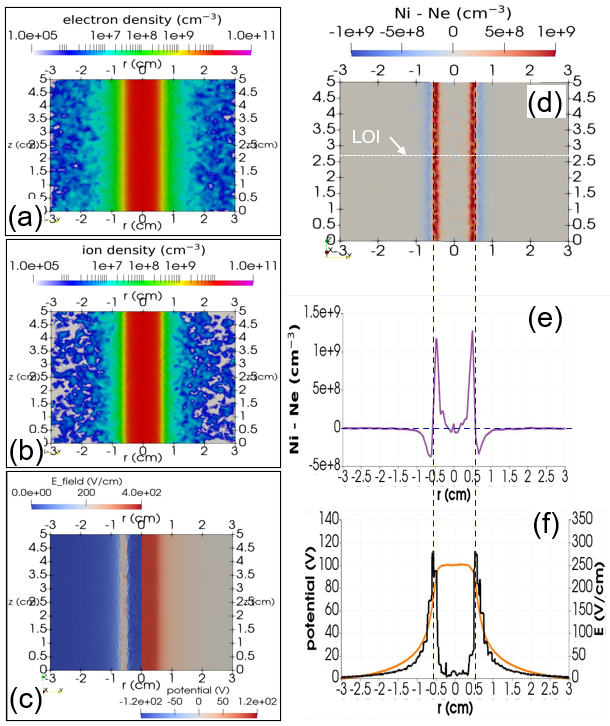}
\caption{Spatial distributions of (a) electron density ($N_e$), (b) ion density ($N_i$) during EUV exposure ($E_b$ = 1mJ) at 40 ns with two boundaries between exposed and unexposed area (c) Spatial distribution of electrostatic potential and electric field at one of the two boundaries mentioned before (d) The spatial distribution of delta ($N_i - N_e$) density which clearly shows DL formation at the boundary. The DL characteristics along the Line of Interest (LOI) are shown as (e) Delta density ($N_i - N_e$) variation and (f) the variation of electrostatic potential and electric field} 
\label{DL_EUV_beam}  
\end{figure}
Theoretically, DL arises as nonlinear solutions to the Vlasov–Poisson system or reduced fluid models, maintained by a balance between particle trapping, current continuity, dissipation and non-Maxwellian velocity distributions which are often linked to electron holes as well as other phase-space structures~\cite{Schamel1972,Newman2001PRL,Singh2011,Torven1999,Forslund1967,Okuda1973,Joyce1978,Newman2001}. The stability of quasi-static DL structures depends on two specific mechanisms: charge separation and pressure balance. Depending on the resulting potential drop, a key distinction is made between ``weak'' ($v_d < v_{th}$) and ``strong'' ($v_d > v_{th}$) DLs where $v_{th}$ and $v_d$ are particle thermal speed and drift speeds respectively. In this context, the Particle‑in‑cell (PIC) simulations have been central in establishing DL as robust, self‑consistent solution of the kinetic plasma equations, free from assumptions about distribution functions or closure relations~\cite{Forslund1967,Okuda1973,Joyce1978,Newman2001,Goldman2003,Decyk2003,Lapenta2010,Divin2012}. 
These simulations reveal that DL acts as localized electrostatic accelerators, efficiently converting field energy into directed particle kinetic energy while regulating current flow. PIC results further indicate that double‑layer stability and strength depend sensitively on boundary conditions, mass ratios, and dimensionality, raising open questions regarding their formation thresholds and interaction with turbulence. {\it The goal of this work is to explore a new parameter regime associated with fast transient DL structural development at the interface of EUV exposed and un-exposed regions at high EUV power using 3DPIC.}

\begin{figure}[h]
\includegraphics[width=0.95\linewidth]{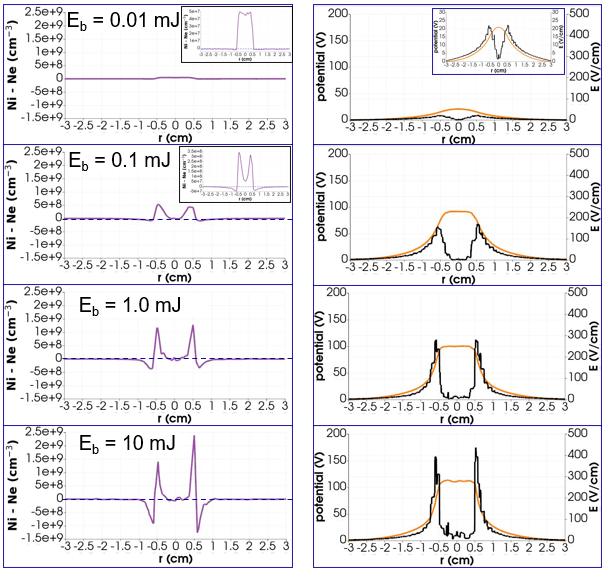}
\caption{The variation of delta ($N_i - N_e$) density within DL for different EUV beam energy is shown in the left panel. The corresponding variations of electrostatic potential and electric field within DL for different EUV beam energy are shown in the right panel. The inset shows plots with zoom-in scales.} 
\label{DL_EUV_beam_energy_variation}  
\end{figure}

The kinetic approach has been adopted to address non-Maxwellian electron energy distribution function (EEDF) by solving Boltzmann-Vlasov equations for relevant plasma species:

\begin{equation}
\frac{\partial f_a}{\partial t} + {\bf v}\nabla_{\bf r}f_a + \frac{q_a{\bf E}_a}{m_a}\nabla_{\bf v}f_a = St(f) 
\end{equation}
coupled with Poisson equation:
\begin{equation}
{\bf \nabla}.{\bf E} = -4\pi\rho  
\end{equation}

Here $f_a$ is the distribution function for speciaes $a$ (= electron/ion), $St(f)$ is the collision integral with Monte Carlo collision scheme (MCC). ${\bf E}$ is the electric field which is coupled with electrostatic potential ($\phi$) as ${\bf E} = -\nabla\phi$. The model follows explicit in time energy conserving formulation \cite{Langdon.1973.energy-conserving}. One of the favorable property of this formulation is that only plasma frequency need to be resolved accurately and the cell size can be much larger than the Debye length \cite{Powis.2024.accuracy}.  That allows to model cooling of the electrons due to collisions with H$_2$ and avoid the need to have too fine mesh in 3D. A brief overview of the model is presented here and further details are reported in prior works \cite{Ast15,Ast16_2}.
The code rely on the unstructured tetrahedral mesh to represent geometry. The tetrahedral shape is chosen as base element due to its robustness within available meshes for the cases of complex geometries. The Poisson equation is solved by the finite element method with linear base elements.  This choice in combination with energy conserving formulation leads to the constant E-field in the cells. That follows from the fact that spatial derivatives of the linear base functions for the potential are constants. To counteract large numerical noise induced by cell-constant E-field a sub stepping scheme in time domain is used. The code checks, if particle crosses the cell boundaries on each time step. If no cell-crossing is detected, then the equation of motion of such particle is computed with standard Verlet integrator. If the particle crosses the cell boundary, than the motion of the particle is computed according to the time spend in each cells. That fractional times are saved for each particle that crosses cell boundaries and used in the velocity update part of the Verlet integrator. That way there is a smooth transition between fast particles, that may cross many cells in one step and slow particles, that stay in the same cell for many time steps. The collisions are sampled for each particle on each time step with help of the Nanbu scheme \cite{Nanbu.1994.simple}. The cross-section set used to compute collisions probabilities includes collisions of electrons (e) and ions with H$_2$. Due to low degree of dissociation (expected to be less than $10^{-4}$), the collisions of electrons and ions with atomic hydrogen are not included in the set. The cross-section set is for electrons and H$_2$ is structured similarly to \cite{Mokrov.2008.monte_carlo}. It includes differential cross-section data from \cite{Brunger.2002.electronmolecule} for excitations and from \cite{Shyn.1981.doubly, Rudd.1993.doubly} for ionization.  
Part of the cross-section set for H$^+$, H$_3^+ $ interaction with H$_2$ is based on the \cite{Simko.1997.transport} and \cite{Peko.1997.total}. The cross-section set and angular dependent sampling routines are validated via simulation of the swarm experiments. The good agreement is found for electron drift velocity, ionization coefficient and for H$^+$ , H$_3^+$ mobility and diffusion in H$_2$. This data set is used to model EUV/DUV induced plasmas~\cite{Ast15,Ast16_2}. And it is further utilized in a separate electromagnetic PIC model of electron cyclotron discharges, with benchmarking performed using a range of probe diagnostics \cite{Ere25} The photo absorption of the EUV radiation in simulations domain is introduced as electrons-ion pairs added to volume. The number of such pairs are added on each time step when the EUV pulse intensity is greater then zero. Their number and positions are sampled according to spatio-temporal and spectral characteristics of the EUV pulse. 

\begin{figure}[h]
\includegraphics[width=0.95\linewidth]{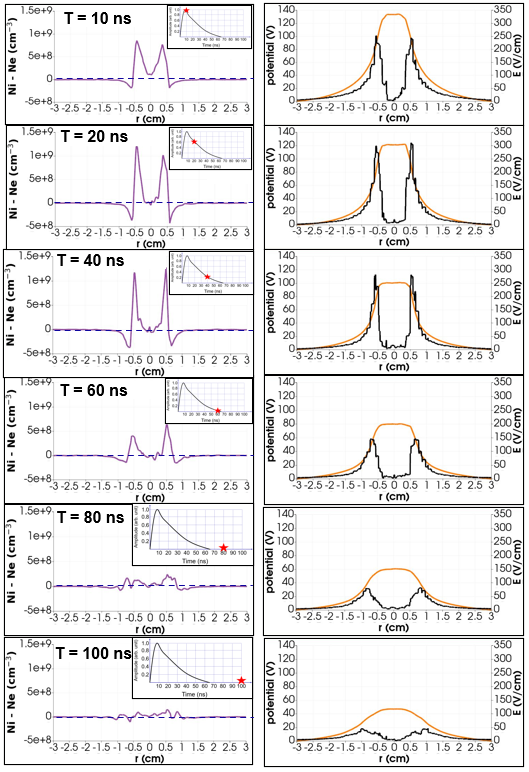}
\caption{The temporal variation of DL for EUV beam energy = 1 mJ. The delta ($N_i - N_e$) density within DL is shown in the left panel. The temporal variations of electrostatic potential and electric field within DL are shown in the left panel. The inset shows the temporal variation of EUV beam. The red star in the inset shows the exact time step when DL charctaristics are plotted. The EUV-ON time scale is $\sim$ 70 ns as shown in  Figure~\ref{EUV_plasma}.} 
\label{DL_EUV_beam_transient_variation}  
\end{figure}

\begin{figure}[h]
\includegraphics[width=0.95\linewidth]{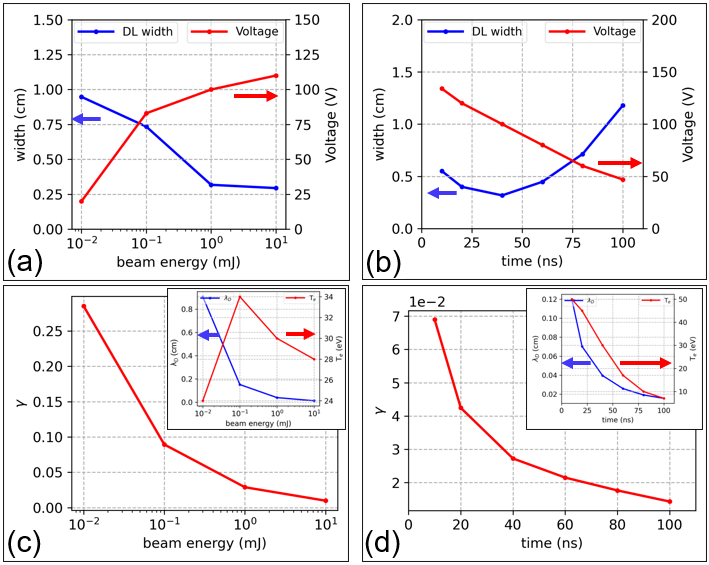}
\caption{The variations of (a) DL width, amplitude and (c) characteristic $\gamma$ parameter are shown with variable EUV beam energy at fixed EUV exposure time (= 40 ns). The variations of (b) DL width, amplitude and (d) characteristic $\gamma$ parameter are shown with transient EUV exposure time for a fixed EUV beam energy (= 1 mJ). Inset shows the associated variation of Debye length ($\lambda_{\rm D}$) and electron temperature ($T_e$). } 
\label{DL_characteristics_variation}  
\end{figure}

The cylindrical computational domain is configured in such a way that the walls are far from the EUV beam boundaries. Both electron and ion density profiles clearly shows the existence of two boundaries on both sides of EUV beam when exposure starts as shown in Figure~\ref{DL_EUV_beam}(a-b). The resulting electrostatic potential and electric field profiles at one of the two boundaries are shown in Figure~\ref{DL_EUV_beam}c. To verify that the EUV beam boundaries are indeed forming DL, the three primary characteristics of DL have been monitored at the boundaries at T = 40 ns: (1) delta ($N_i - N_e$) density maps (2) electrostatic potential profile and (3) electric field profile. The delta density map clearly shows the two populations of electron and ions within DL. To explore the DL quantitatively, all three characteristics mentioned above have been evaluated along the line of interest (LOI) as shown in Figure~\ref{DL_EUV_beam}d. The variation of delta density along the LOI is shown in Figure~\ref{DL_EUV_beam}e. Far from the boundary, the electron and ion densities are equal. As the spontaneously developed DL boundary at the EUV beam edge is approached from the wall side, the higher population of electrons is encountered first followed by ion population. Both these oppositely charged populations are clearly separated within a DL length scale where plasma quasineutrality is violated. In this length scale, the potential typically follows a sigmoid-like curve jumping from a low-potential to high-potential regime. This potential jump is the defining characteristics of a DL. The associated electric field is localized entirely within the transition region with a peak value at the center where the potential gradient is steepest. The FWHM of E-field is defined as DL width. 

The impact of EUV beam energy on DL characteristics has been monitored in Figure~\ref{DL_EUV_beam_energy_variation} where different EUV beam energies ($E_b$ = 0.01, 0.1, 1.0 and 10 mJ) at T = 40 ns are considered.  For very low EUV beam energy ($E_b$ = 0.01 mJ), no such DL is observed. With increasing EUV beam energy, the DL slowly develops and the DL characteristics shows nonlinear growth. The DL width decreases from 5.6 mm ($E_b$ = 0.1 mJ) to 2.9 mm ($E_b$ = 10 mJ) and associated voltage drop increases from 21 V to 108 V as shown in Figure~\ref{DL_characteristics_variation}a. The transient DL development has also been observed for higher EUV beam energy as shown in Figure~\ref{DL_EUV_beam_transient_variation}. The delta ($N_i - N_e$) density variation along with electrostatic potential and electric field within the DL at each step are shown. A clear correlation is observed between EUV exposure profile and DL characteristics. The EUV exposure profile reaches its maximum value at 10 ns and then monotonically decreases towards zero at 70 ns when it is switched off. To capture the complete transient DL behavior, its characteristics have been monitored during EUV-ON period (10, 20, 40 and 60 ns) as well as during EUV-OFF period (80 ns and 100 ns). The DL potential reaches maximum value 120 V at 20 ns and then starts decreasing at subsequent  time steps. During EUV-OFF period, the delta density almost disappears but a residual electrostatic potential and electric field is left over. The variations of electrostatic potential and electric field are shown in Figure~\ref{DL_characteristics_variation}b. The variations of normalized DL parameter, $\gamma [=(\lambda_{\rm D}/L)(e\phi_{\rm D}/T_e)]$ are shown in Figure~\ref{DL_characteristics_variation}(c-d). Such variations of $\gamma$ values are in the range [0.01 - 0.3]. Such $\gamma$ values are comparable with earlier reported values for DL experiments~\cite{Torven1971,Quon1976,Coakley1978} and simulations~\cite{Joyce1978}. With increasing EUV beam energy, the weak DL transforms to a strong DL. However, the strong DL is always associated with EUV-ON time period in the transient process.

\begin{figure}[h]
\includegraphics[width=0.95\linewidth]{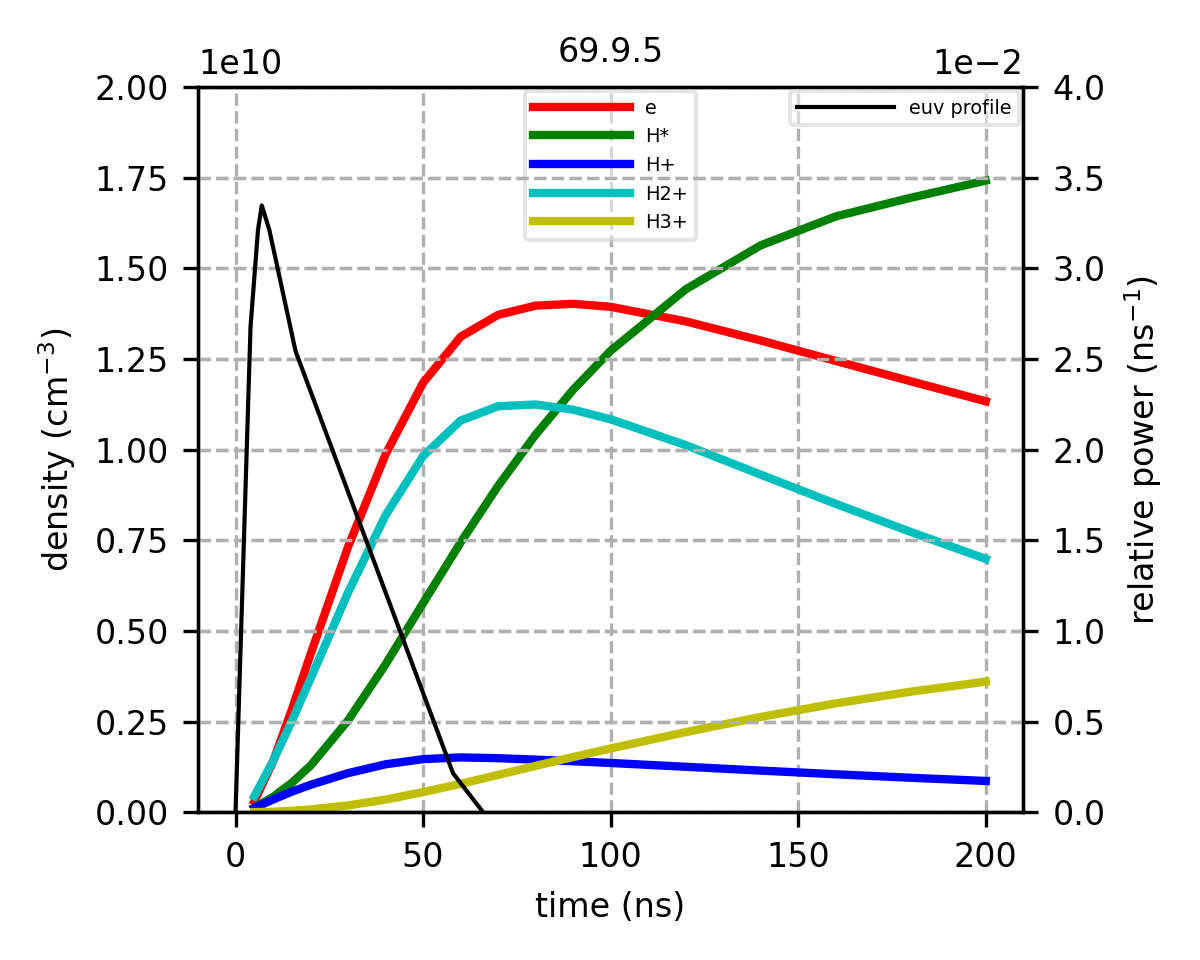}
\caption{The temporal evolution of densities for different EUV induced plasma species inside EUV beam area during DL developments. The dimensionless temporal profile of the euv beam is also shown in /ns units. Its integral equals to 1. The actual temporal power of euv beam may be obtained by multiplying this profile with the energy of a euv beam. The same beam profile has been used for all EUV beam energies. The transient DL develops only during EUV-ON time period.}  
\label{EUV_plasma} 
\end{figure}

During exposure, the EUV photons (energy, $h\nu$ = 92 eV) are abosorved by the background molecular hydrogen gas ($H_2$) and three different processes can trigger simultaneously: photoexcitation, single-/double-photoionization and photodissociation as shown by Equations~\eqref{eq:photoexcitation} -~\eqref{eq:photodissociation}~\cite{Bec19_1,Bec19_2}. Typically, the single photoionization contributes maximum ($\sim 80\%$) of the total ionization event followed by dissociative photoionization ($\sim 15\%$) and double photoionization ($\sim 5\%$). In this time scale primarily the $H_2^+$ ions will be created. Due to dissociative photoionization, the hydrogen molecules get excited as well as ionized. Almost all excited states of $H_2^+$ ions are dissociative and they transform into a hydrogen ion $H^+$ and a radical (excited H atom, $H^*$). The excited state of hydrogen radical depends on the excitation level of $H_2^+$: the ground state radical originates from $H_2^+ (2p\sigma_u)$ whereas $H_2^+ (2p\pi_u)$ generates $H (n=2)$ radicals as per conventions of molecular orbital theory on quantum states. Bulk of the electrons (e) which are generated by single photoionization has a kinetic energy of 77 eV while those created by double photo ionization have a total energy of 60 eV. The electrons which are created by dissociative photoionization have an energy within the range 60 - 77 eV (depending on excited state)
\begin{equation} \label{eq:photoexcitation}
H_2 + h\nu \rightarrow e + H_2^+ 
\end{equation}
\begin{equation} \label{eq:photoionization}
H_2 + h\nu \rightarrow e + H_2^{+*} \rightarrow e + H^* + H^+ 
\end{equation}
\begin{equation} \label{eq:photodissociation}
H_2 + h\nu \rightarrow 2e + H_2^{2+} \rightarrow 2e + 2H^+ 
\end{equation}

At the initial stage of EUV exposure (until 60 ns), the generated electron energy is much smaller than DL potential barrier height (strong DL). As a consequence all these electrons will be trapped between two DLs at two boundaries as they do not have sufficient energy to overcome DL potential barrier. After 60 ns, the height of the DL potential barrier decreases below electron energy (weak DL) so that trapped electrons with higher energy starts to escape from the EUV illuminated zone. At the end of 100 ns, the DL almost disappears so that all the electrons now reach the wall leaving behind heavy and slow ions. Such charge imbalance creates a potential drop between the expanding plasma and wall. Such potential difference will decelerate left-over electrons and accelerate the ions towards walls. At this time scale, the $H_2^+$ ions start to convert stable $H_3^+$ ions:    
\begin{equation} 
H_2 + H_2^+ \rightarrow H_3^{+} + H 
\end{equation}

\begin{figure}[h]
\includegraphics[width=0.95\linewidth]{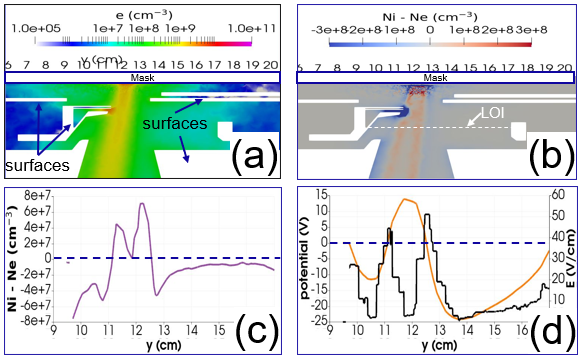}
\caption{The impact of complex geometry (ex. Reticle mini environment with multiple surfaces) on DL chactecteristics due to the presence of wall induced sheath boundaries. (a) Spatial electron density profile during EUV exposure (@ 41 ns) with two boundaries between exposed and unexposed area (b) The spatial delta ($N_i - N_e$) density profile which shows DL formation at the boundary of EUV exposed and unexposed area (c) The delta ($N_i - N_e$) density variation along the line of interest (d) The variation of electrostatic potential and electric field within DL along the line of interest}  
\label{DL_EUV_beam_RME} 
\end{figure}

It is to be noted that so far the DL characteristics have been explored in a simplified cylindrical geometry. However, in case of complex geometry the DL characteristics changes significantly. One such scenario is considered as reticle mini environment (RME) within EUV lithography scanner~\cite{Kerkhof_2021a,chaudhuri_2025} where the walls and multiple hardware surfaces are close to the EUV boundaries as shown in Figure~\ref{DL_EUV_beam_RME}a. The spatial distributio map of delta density (Figure~\ref{DL_EUV_beam_RME}b) as well as its profile on the LOI (Figure~\ref{DL_EUV_beam_RME}c) have been explored. Unlike the simplified cylindrical geometry discussed before, a clear difference is observed in this case: the delta density does not reach smoothly to the zero value on the side where walls associated with hardware surfaces are much closer to the EUV boundary. The associated variation of electrostatic potential and electric field on the LOI are shown in Figure~\ref{DL_EUV_beam_RME}d.

In conclusion, the existence of DL at the boundary between EUV exposed and un-exposed regions have been discovered for the first time using 3DPIC simulation. The typical DL characteristics namely delta density ($N_i - N_e$), electrostatic potential and electric field variations becomes distinct with increasing EUV beam energy. The transient evolution of DL shows correlation between DL characteristics with EUV beam energy profile. The DL characteristics are visible only during EUV-ON scenario. As the EUV is switched off, the DL also dies down quickly. It is also found that complex geometry with multiple hardware surfaces has a strong influence on DL characteristics. At high EUV power, the DL can also trigger localized turbulences close to its boundaries through energy cascade processes. The presence of such DL in high EUV exposure can strongly influence ion acceleration / energy disssipation in complex geometry which can impact nearby hardware surfaces. Also the presence of high electric field within DL can trigger hydrogen radical generation through homolytic cleavage by distorting molecular orbitals, lowering effetive H-H bond dissociation energy as well as making effective coupling between vibrational and thermal excitation. It would be important to execute relevant dedicated experiments to explore detailed physics associated with this phenomena. However, all such detailed investiggations are out of scope for present studies and kept for future investigations.

\end{document}